\documentclass[a4paper,11pt]{article}

\usepackage[T1]{fontenc}
\usepackage[ansinew]{inputenc}
\usepackage{fancyhdr}
\usepackage{amstext}
\usepackage{epsfig}
\usepackage{graphicx}
\usepackage{latexsym}
\usepackage{dcolumn}
\usepackage{bm}
\usepackage{rotating}

\newcommand{\pbar}{$\overline{\text p}$}

\begin{document}

\title{Low-energy antiprotons physics and the FLAIR facility}

\author{E. Widmann\footnote{email: eberhard.widmann@oeaw.ac.at}\\
Stefan Meyer Institut für subatomare Physik, Austrian Academy of
Sciences,\\ Boltzmanngasse 3, A-1090 Vienna, Austria}

\maketitle

\begin{abstract}
FLAIR, the Facility for Low-energy Antiproton and Ion Research has been proposed in 2004 as an extension of the planned FAIR facility at Darmstadt, Germany. FLAIR was not included into the Modularized Start Version of FAIR, but the recent installation of the CRYRING storage ring at GSI Darmstadt has opened new perspectives for physics with low-energy antiprotons at FAIR.
\end{abstract}


Keywords: low-energy antiproton physics, FAIR - Facility for Antiproton and Ion Research, FLAIR - Facility for Low-energy Antiproton and Ion Research

\section{Introduction}

Physics with low-energy antiprotons is currently ongoing at the Antiproton Decelerator of CERN \cite{Maury:1997la,Eriksson:2013aa}, which started operation in the year 2000. In 2004, the low-energy antiproton community launched a new initiative for a next-generation facility called Facility for Low-energy Antiproton and Ion Research FLAIR \cite{FLAIR-LOI,Widmann:2005ys} at the FAIR facility that was planned to be built in Darmstadt. At that time the long-term future of CERN-AD and thus of the field of low-energy antiproton physics was uncertain, and FAIR was the only other facility planned where high-intensity cooled antiproton beams would be available. The LOI of FLAIR was approved and a Baseline Technical Report \cite{FLAIR-TP} was submitted and approved 2005. Thus FLAIR was included in the full FAIR facility \cite{BTR} in 2009. 

In the following it turned out that it was much more difficult and time consuming to establish the international FAIR facility, and for the formal foundation of FAIR in 2010 the original program had to be reduced to what is called the Modularized Start Version MSV. FLAIR is not part of the MSV and was deferred to a later phase of FAIR. In the mean time, however, the successes of the physics program at the AD had convinced CERN to extend the antiproton program and a further acceleration ring called ELENA \cite{Maury:2014aa} which closely resembles the LSR ring of FLAIR was approved in 2011. It is currently under construction and will go into operation in 2017. 

In spite of not being included in the MSV, a development started by the FLAIR collaboration led to the transfer of the CRYRING storage ring from Stockholm University to GSI in 2012. CRYRING, which was chosen by FLAIR to be used as its central storage ring LSR, is now installed at GSI behind the existing ESR storage ring and is being commissioned to use highly charged ions from ESR as well as from an ion source. If a way can be found to bring antiprotons from the production target to CRYRING, physics with low-energy antiprotons can start there. Since CRYRING was designed to provide continuous beams that are not available at CERN, nuclear and particle physics type experiments could be uniquely  performed there.

\section{The FLAIR facility as proposed in 2005}

\begin{figure}[t]
\begin{center}
\epsfig{file=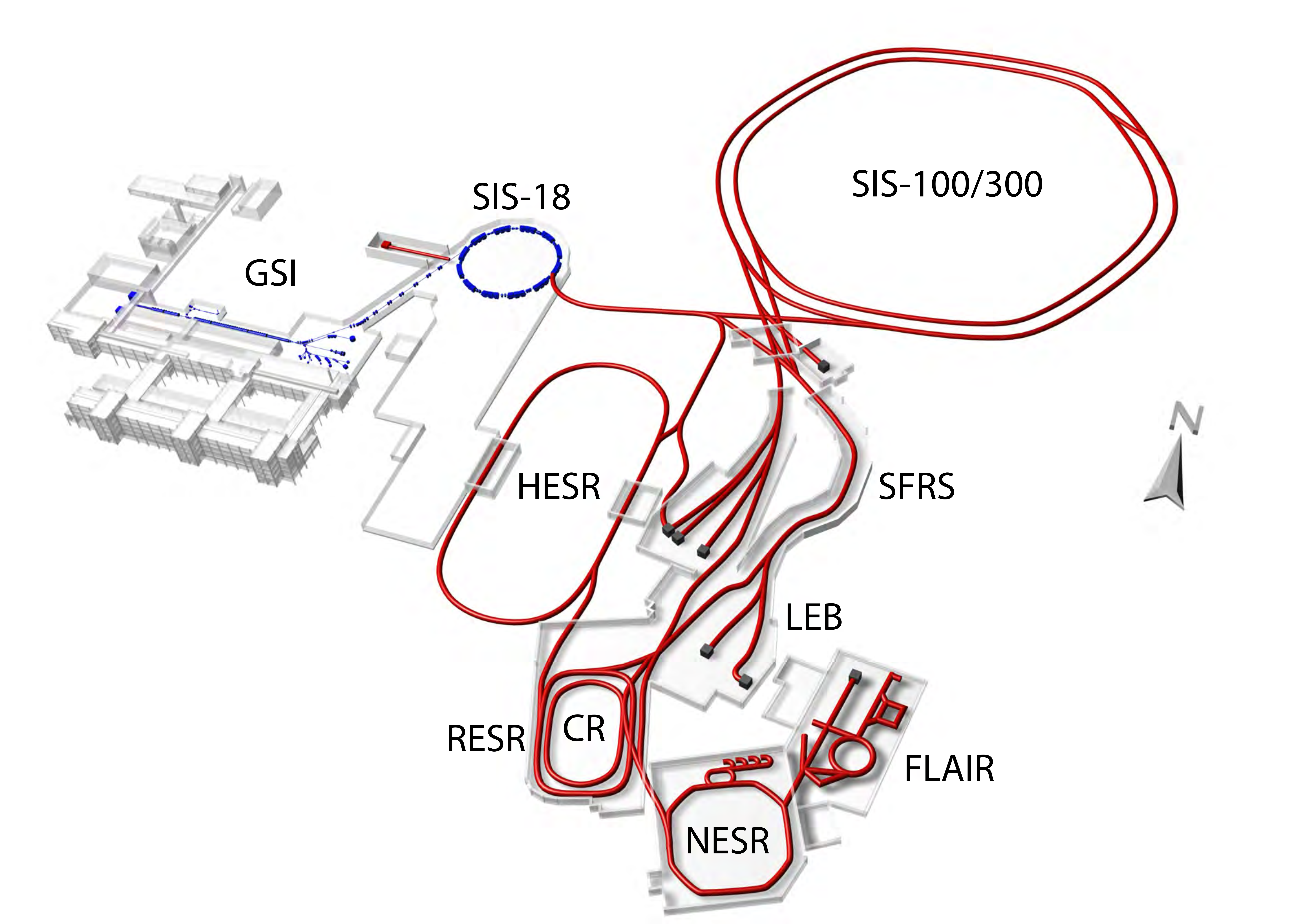,width=0.6\textwidth}
\caption{Overview of the FAIR facility. Left the existing GSI
facility is shown with the heavy ion synchrotron SIS-18, and right the newly planned facilities. Bold
lines symbolize storage rings and beam pipes. SIS-100/300: heavy
ion synchrotron with 100/300 Tm magnetic rigidity. SFRS: super
fragment separator for studying short-lived nuclei. LEB: low energy branch of the SFRS. CR: collector
ring for antiprotons. RESR: accumulation ring for antiprotons.
HESR: high-energy storage ring for antiprotons. NESR: new
experimental storage ring for experiments with highly charged ions
and to decelerate antiprotons, and attached the FLAIR facility.} \label{fig:FAIR}
\end{center}
\end{figure}

\begin{figure}[t]
\begin{center}
\epsfig{file=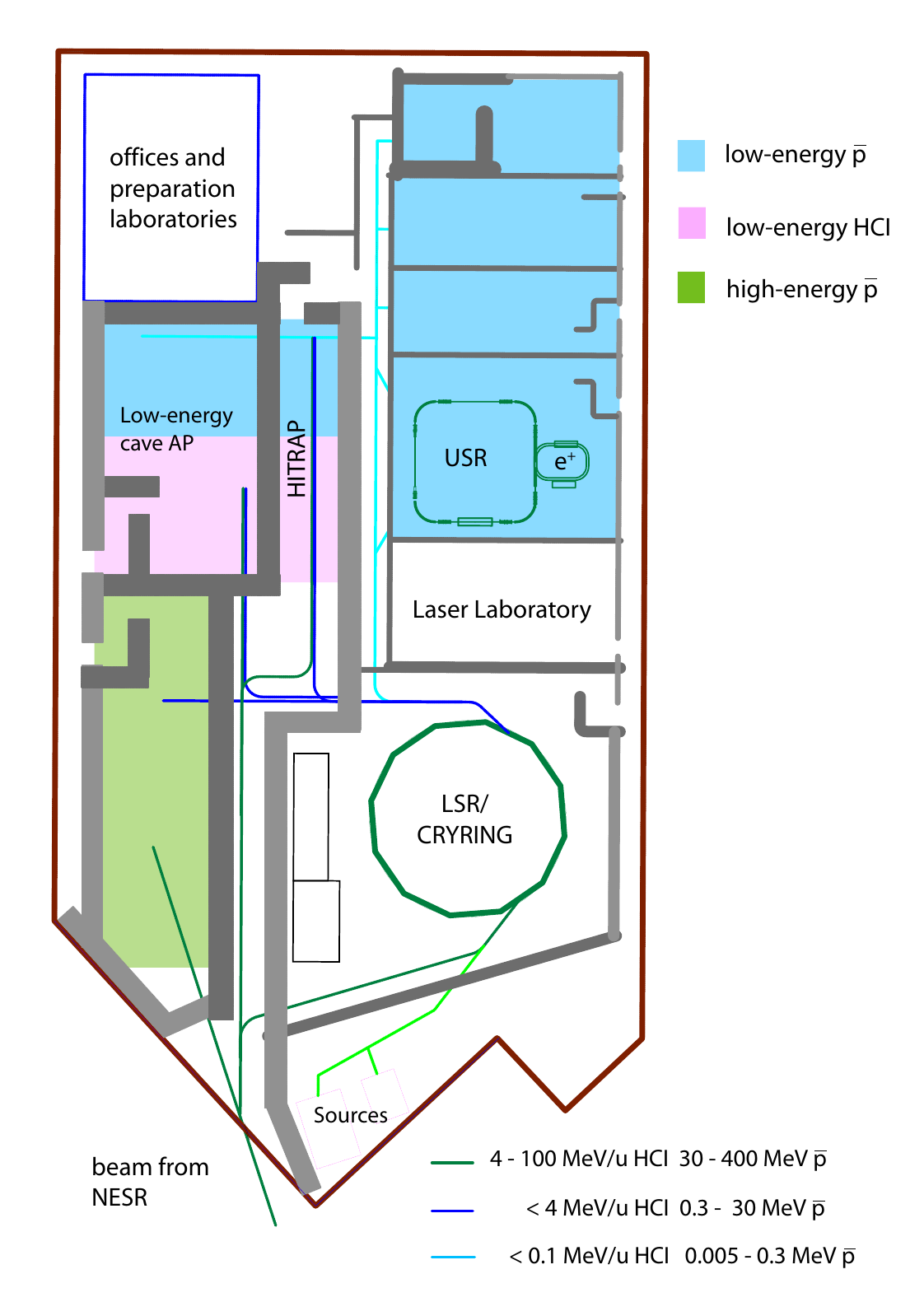,width=0.4\textwidth}
\caption{Layout of the FLAIR hall from the BTR. The colours indicate available energies of antiprotons as detailed in the legend.} \label{fig:FLAIR}
\end{center}
\end{figure}

Fig.~\ref{fig:FAIR} shows the full FAIR facility as described in its Baseline Technical Report (BTR) \cite{BTR}. Antiprotons from the CR/RESR were going to be decelerated to 30 MeV before beeing transferred into the FLAIR hall. The layout shown in Fig.~\ref{fig:FLAIR} was developed by the FLAIR collaboration for the FLAIR BTR  \cite{FLAIR-TP}  with the aim of designing a next generation facility that would produce cooled antiproton beams with higher rate  (making use of the accumulation in the RESR) at lower energies than available at that time, and with both fast and slow extraction. A detailed estimation showed that FLAIR could produce a factor 100 more antiprotons per second trapped in Penning traps or stopped in low-density gas, and continuous beams of up to $10^6$ \pbar/s  \cite{FLAIR-LOI,Widmann:2005ys}. The available rates in storage rings at these low energies are limited by the space charge, so that the total \pbar\ consumption of FLAIR is  limited to about 10\% of antiprotons produced at FAIR. One of the key features of FLAIR is that all components can be used with antiprotons as well as highly charged ions.

To achieve these goals the FLAIR facility plans to use two storage rings, a magnetic one (LSR) and an electrostatic one (USR), and the Penning trap system HITRAP \cite{Kluge:2008fk}, cf. Fig.~\ref{fig:FLAIR}. The LSR would receive the beams from NESR, decelerate them and either directly extract them for experiments or deliver them to USR or HITRAP for further deceleration. In addition to these decelerated beams, direct beams from NESR for higher energies were foreseen to be delivered to two areas. The available beam energies in the various areas can be seen in Fig.~\ref{fig:FLAIR}.

For LSR the CRYRING accelerator from Manne Siegbahn Laboratory of Stockholm University, Sweden, was chosen by the FLAIR collaboration in 2007. The USR is being developed by C.P. Welsch at Cockcroft Institute, UK \cite{Welsch:2012aa}. At MPI-K Heidelberg the Cryogenic Storage Ring CSR \cite{Hahn:2011fk} is under construction which serves as a prototype for the USR. HITRAP is currently being installed at the ESR storage ring of GSI, so that all key components of FLAIR are well on track.


The main focus of the physics program as presented in the FLAIR letter of intent \cite{FLAIR-LOI,Widmann:2005ys} is the formation and spectroscopy of antihydrogen and the spectroscopy of antiprotonic atoms. These topics are being pursued  at CERN-AD and significant progress has been made, from the formation of slow antihydrogen atoms from nested Penning traps by the ATHENA \cite{ATHENA:2002} and ATRAP \cite{ATRAP:Hbar:2002} collaborations to the first trapping of antihydrogen atoms by ALPHA \cite{Andresen:2010jba} and ATRAP \cite{Gabrielse:2012aa} in recent years. ASACUSA has been continuing precision spectroscopy of antiprotonic helium to ppb precision \cite{Hori:2011aa} and prepared a beam of antihydrogen atoms for hyperfine spectroscopy \cite{Kuroda:2014aa}. The AEgIS collaboration has started, aiming at a  precise measurement of the gravitational interaction of antimatter and recently published a proof-of-principle measurement of the deflection of antiprotons in a Moiré reflectometer \cite{Aghion:2014aa}. The BASE collaboration aims at measuring the magnetic moment of the proton and antiproton to highest precision and has recently reached a precision of 3 ppb for the proton \cite{Mooser:2014aa}. A further experiment pursuing  a gravity measurement, GBAR \cite{Perez:2012aa}, has been approved and will start taking beam after the completion of ELENA. These experiments are now slowly progressing towards spectroscopy and gravity measurements and will continue to operate at CERN for one decade or more until they reach significant precision for the tests of CPT symmetry or gravity.

The physics program of FLAIR using antiprotons as hadronic probes, which was also described in detail in the letter of intent, can not be done at CERN-AD even with ELENA because of the lack of slow extracted beam. These experiments are still waiting for a facility providing continuous antiproton beams.

\section{Opportunities for low-energy antiprotons in the FAIR MSV}

In the MSV of FAIR neither the FLAIR facility nor the NESR ring (c. Fig.~\ref{fig:FAIR}), which is needed to decelerate antiprotons for injection into the LSR, is included, thus pushing FLAIR into a future phase 2 of FAIR.  Experiments with fast extracted antiprotons can continue at CERN where due to the approval of ELENA the low-energy antiproton physics program has now a medium term future. 

A new perspective emerged in 2012, when as a consequence of the activities of the FLAIR and SPARC collaborations the CRYRING storage ring was transferred from MSL Stockholm to GSI, and is now being set up behind the ESR, 
where it can receive heavy ion beams from the GSI SIS-18 either directly or through the Fragment Separator FRS \cite{Herfurth:2013aa}. The current physics program is focussed on the use of highly charged ions from the ESR and thus part of the program of the SPARC collaboration \cite{Stohlker:2014aa}.  It was, however, early noticed (see e.g. \cite{Widmann:2014aa}) that  the sequence ESR - CRYRING could replace NESR - CRYRING as a first step of realising FLAIR in the ESR hall of GSI, if antiproton beams could be brought from the CR to ESR. 

Such possibilities were studied in terms of technical feasibility. Apart from the technical question of how to build a beam line from the CR back to the exiting GSI premises, some problems have to be solved. The ESR has a magnetic rigidity of 10 Tm while the CR is planned to be built with a fixed rigidity of 13 Tm. Several ideas have been tried to bridge this gap, with the best one in terms of \pbar\ intensity being the usage of the HESR to decelerate antiproton below the transition energy of the ESR and then to transfer \pbar\ from HESR to ESR. 
This scenario has been studied in detail in terms of accelerator physics in \cite{Katayama:aa}.

Main use of the HESR is to provide cooled antiprotons beams between 1.5 and 15 GeV/$c$ momentum for the $\mathrm{P}$ANDA experiment, for which $10^{10}$ \pbar\ will be needed to be accumulated. In the MSV this has to be done at 3 GeV/$c$ in the HESR because the RESR is not available. Katayama et al.  \cite{Katayama:aa} show that  in this  way $10^{9}$ \pbar\ could be accumulated and decelerated to 1 GeV/$c$ in a cycle time of approx. 200 s, transferred to ESR and decelerated by both ESR and CRYRING in another 20 s. The overall efficiency of the process is 80\%, yielding $8\times10^8$ \pbar\ every 220 s or $3.6\times10^6$ \pbar/s in average in this assumed exclusive use of HESR for producing low-energy antiprotons. 

A second way of operation would be parasitic to PANDA. PANDA will start with $10^{10}$ \pbar\  which will be reduced by a factor 10 in about 1--2 hours, leaving $10^{9}$ \pbar. These could be cooled in about 100 s to 1 GeV/$c$ and transferred to ESR, yielding $10^{9}$ \pbar\ every $\approx 5000$ s corresponding to $2\times10^5$ \pbar/s in average, very similar to the production rate at CERN-AD and the assumed rates in the original FLAIR BTR.

These antiprotons could be provided in pulsed as well as in slow extraction, creating the unique possibility of using continuous beams of low-energy antiprotons for nuclear and particle physics types of experiments. Some of them were already described in the FLAIR LoI \cite{FLAIR-LOI,Widmann:2005ys}: X-rays of light antiprotonic atoms to study the nucleon-antinucleon interaction \cite{Gotta:2004aa}, X-rays of heavy antiprotonic atoms to study the nuclear periphery ({i. e.} the neutron or proton halo structure) \cite{Trzcinska:01}, production of strangeness $-2$ baryons \cite{Winter:2005vn}.

\begin{figure}[h]
\begin{center}
\epsfig{file=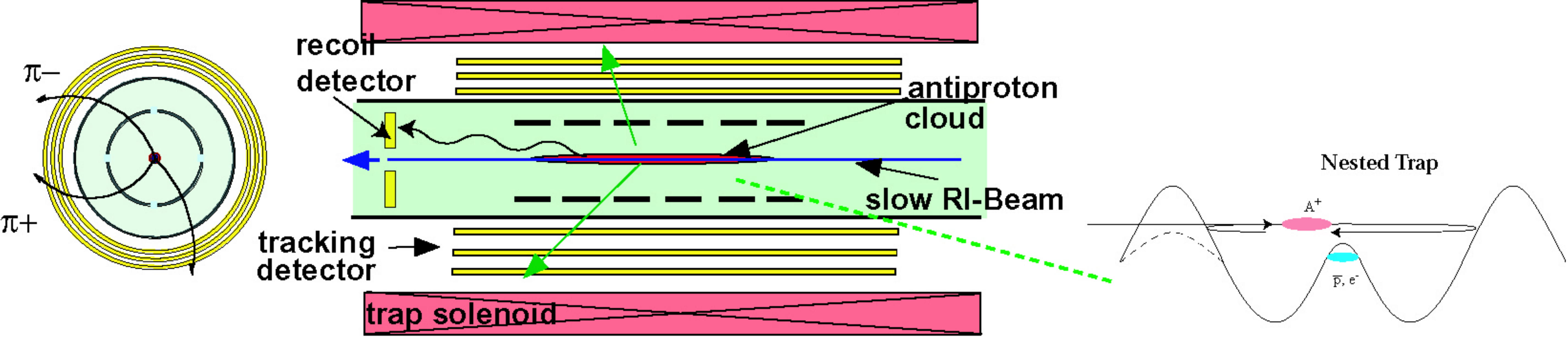,width=0.9\textwidth}
\caption{Schematic setup for the exo+pbar experiment showing a solenoid magnet (center) containing a nested Penning trap (right). One possibility of detection is to track charged particles inside the solenoid (left) and to extract halo factors from the ratio of positive and negative pions produced in the annihilation of the antiproton with a nucleon at the nuclear surface. From \cite{FLAIR-LOI}. \label{fig:exo+pbar}}
\end{center}
\end{figure}

Some of the topics have in the mean time been further developed. The Exo+pbar experiment was proposed \cite{Wada-leap03,Wada:2005aa} which extends the usage of antiprotons as probe of the nuclear periphery to unstable nuclei produced (in analogy to the production of antihydrogen from antiprotons and positrons) in nested Penning traps from short-lived isotopes and antiprotons (cf. Fig.~\ref{fig:exo+pbar}). Antiprotons are one of the few probes to directly measure the neutron radii of nuclei, thus Exo+pbar would be one of the most attractive experiments at a facility being able to provide both low-energy antiprotons and unstable nuclei (which in the MSV is possible through the FRS of GSI). 

\begin{figure}[h]
\begin{center}
\epsfig{file=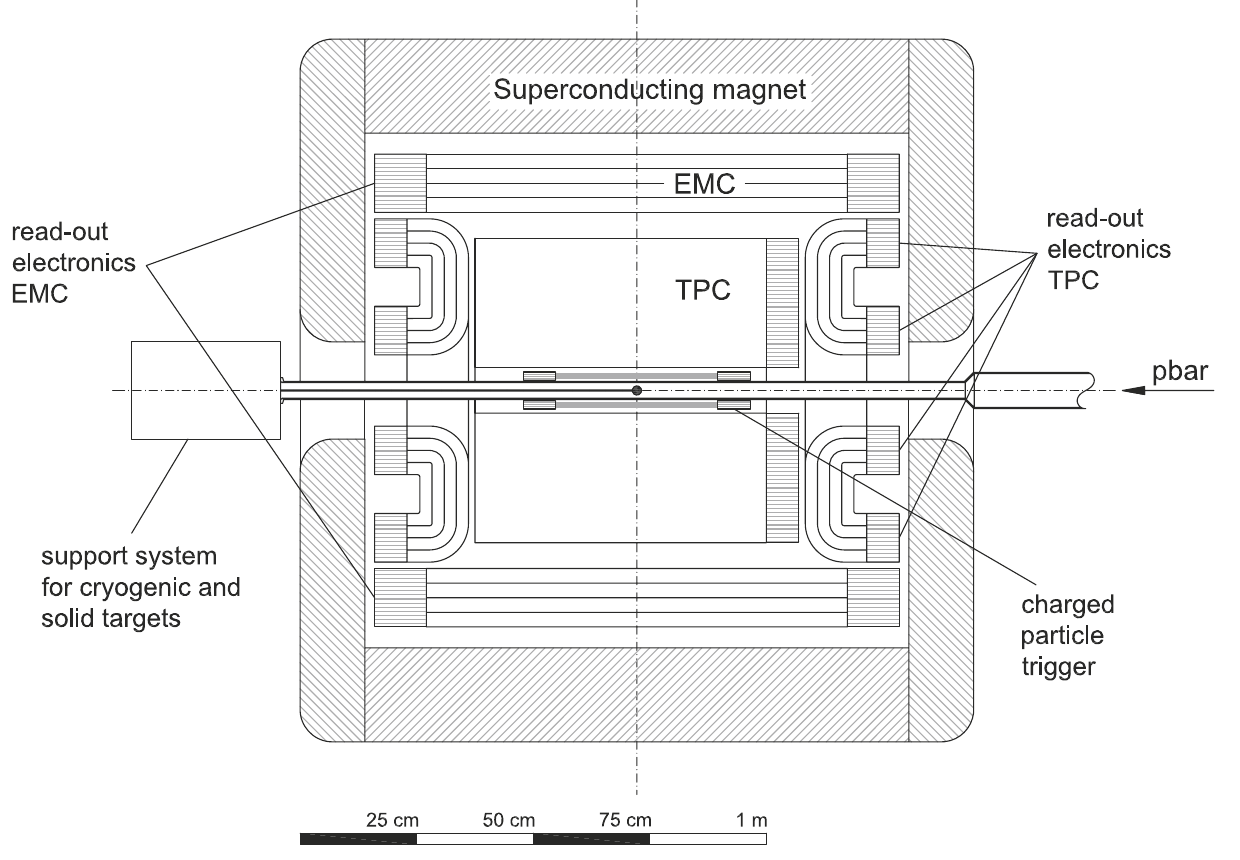,width=0.6\textwidth}
\caption{Schematic layout of a possible 4$\pi$ detector for use with stopped antiprotons. From \cite{Zmeskal:2009zr}.} \label{fig:strangeness-detector}
\end{center}
\end{figure}

A further subject which showed activities in the last years is the strangeness production with stopped antiprotons, partly because of its potential connection to kaonic nuclear bound states that have been predicted \cite{Akaishi:2002qf} and are being searched for in various experiments using kaon and proton beams. Their production using \pbar p annihilation was suggested \cite{Kienle:2005bh,KIENLE:2007ys}, and subsequently experiments at FLAIR \cite{Zmeskal:2009zr} and for J-PARC using a secondary antiproton beam \cite{Sakuma:2012ve} were discussed. For this purpose a 4$\pi$ detector (cf. Fig.~\ref{fig:strangeness-detector}) with the capability of detecting all charged and neutral decay products, which can be also used to study other hadron physics phenomena, will be needed.

\section{Conclusions}

Low-energy antiproton physics using pulsed beams is continuing at CERN-AD where -- because of the ELENA ring under construction -- the physics program focussed around antihydrogen spectroscopy and gravity measurements will continue for at least the next one or two decades. The FLAIR facility, formally being outside the MSV for FAIR, has now a chance of becoming possible in a first stage within the MSV by the addition of a beam line bringing antiprotons from CR to ESR and making use of CRYRING and the current GSI experimental hall, {\em i. e.} with a modest investment. Experiments uniquely possible with slowly extracted low-energy antiprotons from CRYRING@ESR are studies of the nuclear periphery of unstable nuclei and the strangeness production from stopped antiprotons.

\section*{Acknowledgements}

The author would like to thank the members of the FLAIR collaboration for their continued support, and Drs. Thomas St{\"o}hlker, B. Franzke, Y. Litvinov, and T. Katayama for stimulating discussions on the possibility of providing antiprotons for CRYRING@ESR.



\newcommand{\SortNoop}[1]{}

\end{document}